%% file: direct-rb-arXiv-final.tex
\documentclass[aps,prl,superscriptaddress,twocolumn,showpacs,longbibliography, titlepage=false]{revtex4-1}
\usepackage{amsmath}
\usepackage{amssymb}
\usepackage{subfigure}
\usepackage{natbib}
\usepackage{color}
\usepackage{blkarray}
\usepackage{dsfont}
\usepackage{MnSymbol}
\usepackage{mathrsfs}
\usepackage{bbold}
\usepackage[breaklinks=true,colorlinks=true,linkcolor=blue,urlcolor=blue,citecolor=blue]{hyperref}
\usepackage{bbm}
\usepackage{hyperref}
\usepackage{amsthm}
\usepackage{graphicx}
\usepackage{enumitem}

\graphicspath{{./figures/}}

\def\ket#1{{\left| #1 \right\rangle}}

\begin{document}
\title{Direct randomized benchmarking for multi-qubit devices}
\author{Timothy J. Proctor}
\affiliation{Quantum Performance Laboratory, Sandia National Laboratories, Livermore, CA 94550, USA}
\author{Arnaud Carignan-Dugas}
\affiliation{Institute for Quantum Computing and the Department of Applied Mathematics, University of Waterloo, Waterloo, Ontario N2L 3G1, Canada}
\author{Kenneth Rudinger}
\author{Erik Nielsen}
\author{Robin Blume-Kohout}
\affiliation{Quantum Performance Laboratory, Sandia National Laboratories, Albuquerque, NM 87185, USA}
\author{Kevin Young}
\affiliation{Quantum Performance Laboratory, Sandia National Laboratories, Livermore, CA 94550, USA}
\date{\today}
\begin{abstract}
Benchmarking methods that can be adapted to multi-qubit systems are essential for assessing the overall or ``holistic'' performance of nascent quantum processors. The current industry standard is Clifford randomized benchmarking (RB), which measures a single error rate that quantifies overall performance. But scaling Clifford RB to many qubits is surprisingly hard.  It has only been performed on 1, 2, and 3 qubits as of this writing.  This reflects a fundamental inefficiency in Clifford RB: the $n$-qubit Clifford gates at its core have to be compiled into large circuits over the 1- and 2-qubit gates native to a device. As $n$ grows, the quality of these Clifford gates quickly degrades, making Clifford RB impractical at relatively low $n$. In this Letter, we propose a \emph{direct} RB protocol that mostly avoids compiling.  Instead, it uses random circuits over the native gates in a device, seeded by an initial layer of Clifford-like randomization.  We demonstrate this protocol experimentally on 2 -- 5 qubits, using the publicly available IBMQX5.  We believe this to be the greatest number of qubits holistically benchmarked, and this was achieved on a freely available device without any special tuning up.  Our protocol retains the simplicity and convenient properties of Clifford RB: it estimates an error rate from an exponential decay. But it can be extended to processors with more qubits -- we present simulations on 10+ qubits -- and it reports a more directly informative and flexible error rate than the one reported by Clifford RB.  We show how to use this flexibility to measure separate error rates for distinct sets of gates, which includes tasks such as measuring an average \textsc{cnot} error rate.
\end{abstract}
\maketitle

\noindent With quantum processors incorporating 5 -- 20 qubits now commonplace \cite{ibmqx,ibmqx-backend,linke2017experimental,figgatt2017complete,otterbach2017unsupervised,friis2018observation,riofrio2017experimental,kelly2015state,neill2018blueprint,song201710,fu2017experimental,intel-17q-pressrelease}, and 50+ qubits expected soon \cite{kelly2018engineering,IBM-50q-pressrelease,intel-49q-pressrelease}, efficient, holistic benchmarks are becoming increasingly important. Isolated qubits or coupled pairs can be studied in detail with tomographic methods \cite{merkel2013self,blume2016certifying,greenbaum2015introduction,kimmel2015robust,rudinger2017experimental}, but the required resources scale exponentially with qubit number $n$, making these techniques infeasible for $n \gg 2$ qubits. And while an entire device could be characterized two qubits at a time, this often results in over-optimistic estimates of device performance that ignore crosstalk and collective dephasing effects.  What is needed instead is a family of \emph{holistic} benchmarks that quantify the performance of a device as a whole.  Randomized benchmarking (RB) methods \cite{emerson2005scalable,emerson2007symmetrized,knill2008randomized,magesan2011scalable,magesan2012characterizing,carignan2015characterizing,cross2016scalable,brown2018randomized,hashagen2018real,magesan2011scalable,magesan2012characterizing,carignan2015characterizing,cross2016scalable,brown2018randomized,hashagen2018real} avoid the specific scaling problems that afflict tomography -- in RB, both the number of experiments \cite{helsen2017multiqubit} and the complexity of the data analysis \cite{magesan2012characterizing} are independent of $n$ -- but introduce a new scaling problem in the form of \emph{gate compilation}.

Although a quantum processor's native gates typically include only a few one- and two-qubit operations, the ``gates'' benchmarked by RB are elements of an exponentially large $n$-qubit group 2-design (e.g., the Clifford group).  These gates must be \emph{compiled} into the native gate set \cite{aaronson2004improved,patel2003efficient}. As the number of qubits increases, the circuit depth and infidelity of these compiled group elements grow rapidly, rendering current RB protocols impractical for relatively small $n$, even with state-of-the-art gates. The industry-standard protocol laid out by Magesan \emph{et al.}~\cite{magesan2011scalable,magesan2012characterizing} -- which we will refer to as \emph{Clifford randomized benchmarking} (CRB) -- has been widely used to benchmark \cite{yoneda2018quantum,zajac2018resonantly,watson2018programmable,nichol2017high,veldhorst2014addressable,corcoles2013process,xia2015randomized,corcoles2015demonstration,chen2016measuring,muhonen2015quantifying,barends2014superconducting,raftery2017direct} and calibrate \cite{rol2017restless,kelly2014optimal} both individual qubits and pairs of qubits, but we are aware of just one reported application to three qubits \cite{mckay2017three}, and none to four or more.

\begin{figure}[b!]
\includegraphics[width=8.5cm]{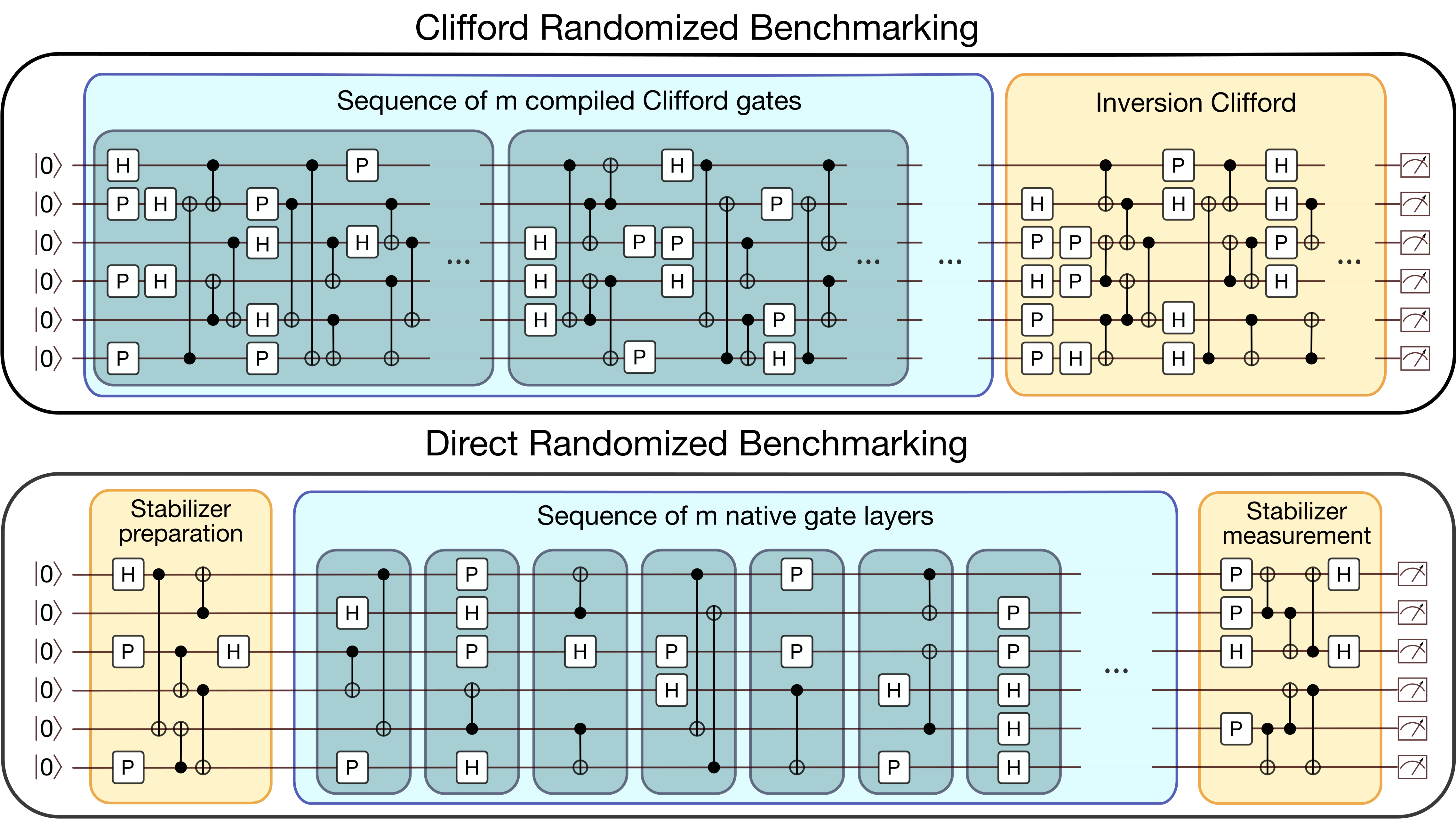}
\caption{A cartoon illustrating the circuits used in Clifford RB and the streamlined \emph{direct} RB protocol that we propose.}
\label{Fig:schematic}
\end{figure}

Another consequence of compilation is that, instead of quantifying native gate performance, CRB measures the error \emph{per compiled group element}.  Although this is sometimes translated into a native gate error rate, e.g., by dividing it by the average circuit size of a compiled Clifford  \cite{muhonen2015quantifying,barends2014superconducting,raftery2017direct}, this is ad hoc and not always reliable \cite{epstein2014investigating}.  Moreover, error rates obtained this way are hard to interpret for $n \gg 1$ CRB, where error rates can vary widely between native gates. 

In this Letter we propose and demonstrate \emph{direct randomized benchmarking} (DRB), an RB protocol that directly benchmarks the native gates of a device.  Like CRB, our DRB protocol utilizes random circuits of variable length, but these circuits consist of the native gates of the device, rather than compiled Clifford operations (see Fig.~\ref{Fig:schematic}). Our protocol is not infinitely scalable, but the simplified structure enables DRB to be successfully implemented on significantly more qubits than CRB. Moreover, DRB preserves the core simplicity of CRB: it estimates an error rate from an exponential decay.

We anticipate that DRB will be an important tool for characterizing current multi-qubit devices. For this reason, this Letter focuses on the practical applications of DRB. We present experiments on 2 -- 5 qubits and simulations on 2 -- 10+ qubits. These examples show that DRB works, demonstrate how our protocol improves on current methods, and show that DRB can be implemented on significantly more than two qubits on current devices. We follow these demonstrations with arguments for why DRB is broadly reliable, but this Letter does not contain a \emph{comprehensive} theory for DRB -- that will be presented in a series of future papers.

\vspace{0.1cm}
\noindent
{\bf Direct randomized benchmarking --} DRB is a protocol to directly benchmark the native gates in a device. There is flexibility in defining a device's ``native gates''. For DRB we only require that they generate the $n$-qubit Clifford group $\mathbb{C}_n$ \footnote{Extensions to other group 2-designs are possible.}. Normally, they will be all the $n$-qubit Clifford operations that can be implemented by depth-1 circuits, e.g., by parallel 1- and 2-qubit gates (see Fig.~\ref{Fig:schematic}). We call these \emph{circuit layers} or ($n$-qubit) \emph{native gates}.

Just as CRB uses sequences of random Cliffords, DRB uses sequences of random circuit layers.  But whereas the Cliffords in CRB are supposed to be uniformly random, DRB allows the circuit layers to be sampled according to a \emph{user-specified} probability distribution $\Omega$. Many distributions are permissible, but, to ensure reliability, $\Omega$ must have support on a subset of the gates that generates $\mathbb{C}_n$ and $\Omega$-random circuits must quickly spread errors (see later).

The $n$-qubit DRB protocol is defined as follows (note that all operations are assumed to be imperfect):
\begin{enumerate}
\item  For a range of lengths $m \geq 0 $, repeat the following $ k_m \gg 1$ times:
\vspace{-0.1cm}
\begin{enumerate}
\renewcommand{\labelenumii}{\labelenumi\arabic{enumii}.}
\item Sample a uniformly random $n$-qubit stabilizer state $\ket\psi$.
\item Sample an $m$-layer circuit $\mathcal{U}_m$, where each layer is drawn independently from some user-specified distribution $\Omega$ over all $n$-qubit native gates.
\item Repeat the following $N \geq 1$ times:
\begin{enumerate}
\item[1.3.1] Initialize the qubits in $\ket{0}^{\otimes n}$.
\item[1.3.2]  Implement a circuit to map $\ket{0}^{\otimes n}\to\ket\psi$.
\item[1.3.3]  Implement the sampled $\mathcal{U}_m$ circuit.
\item[1.3.4]  Implement a circuit that maps $\mathcal{U}_m\ket\psi$ to a known computational basis state $\ket{s}$.
\item[1.3.5]  Measure all $n$ qubits and record whether the outcome is $s$ (success) or not (failure).
\end{enumerate}
\end{enumerate}
\vspace{-0.1cm}
\item Calculate the average probability of success $P_m$ at each length $m$, averaged over the $k_m$ randomly sampled circuits and the $N$ trials for each circuit.
\vspace{-0.1cm}
\item Fit $P_m$ to $P_m = A + Bp^m$, where $A$, $B$ and $p$ are fit parameters.
\vspace{-0.1cm}
\item The $\Omega$-averaged DRB error rate of the native gates is $r = (4^n - 1)(1 - p)/4^n$.
\end{enumerate}
The $n$-dependent rescaling used above is different from that in common usage \cite{knill2008randomized,magesan2011scalable,magesan2012characterizing}. Using our convention, $r$ corresponds to the probability of an error when the errors are stochastic (see later). This is particularly convenient when varying $n$.

DRB is similar to the earliest implementations of RB. Both the 1-qubit RB experiments of Knill \emph{et al}.~\cite{knill2008randomized} and the 3-qubit experiments of Ryan \emph{et al}.~\cite{ryan2009randomized} utilize random sequences of group generators, and so are specific examples of DRB \emph{without} the stabilizer state preparation step and flexible sampling. These additional features, however, are essential to DRB: they make DRB provably reliable under broad conditions, and allow us to separate the error rate into contributions from distinct sets of gates.

\begin{figure}[t!]
\includegraphics[width=0.48\textwidth]{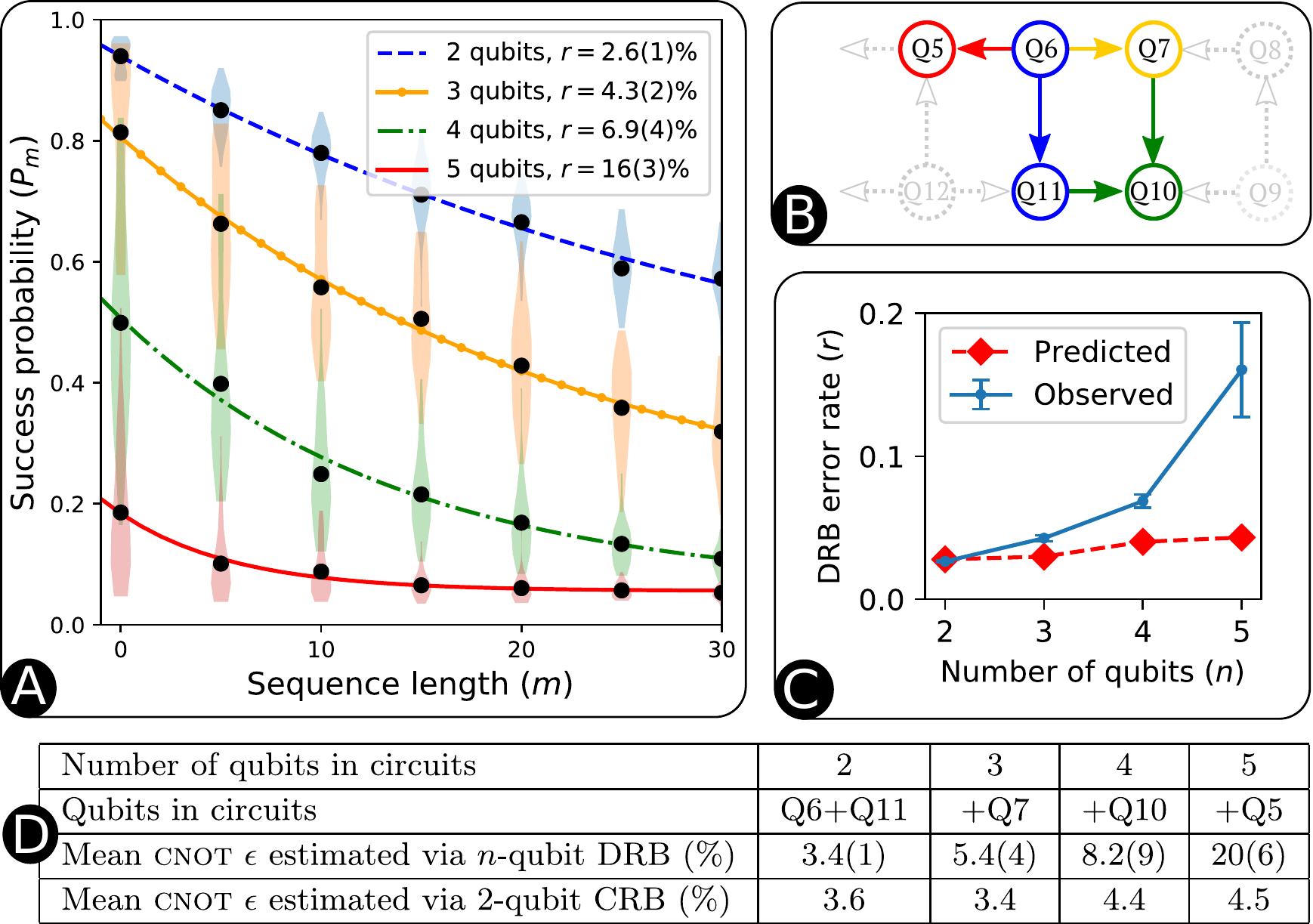}
\caption{Experimental 2 -- 5 qubit DRB on IBMQX5. {\bf A.} Success probability decays. The points are average success probabilities $P_m$, and the violin plots show the distributions of the success probabilities at each length over circuits (there are 28 circuits per length). The curves are obtained from fitting to $P_m = A + Bp^m$, and $r = (4^n-1)(1-p)/4^n$. {\bf B.} A schematic of IBMQX5. The colors match those in A and correspond to the additional qubits/\textsc{cnot}s added from $n \to n+1$ qubit DRB (see also D). {\bf C.} Observed $r$ versus $n$, and predictions from 1- and 2-qubit CRB calibration data. {\bf D.} Estimates of the average \textsc{cnot} error rate in $n$-qubit circuits, obtained by comparing the data in A with additional DRB data that used circuits with fewer \textsc{cnot}s per layer.}
\label{Fig:IBMQX-decays}
\end{figure}

\vspace{0.1cm}
\noindent
{\bf What DRB measures --} To interpret DRB results it is important to understand what DRB measures. Assume that the gate errors are stochastic, which can be enforced to a good approximation by, e.g., Pauli-frame randomization \cite{knill2005quantum,ware2018experimental,wallman2015noise} or by following each layer in DRB with a random $n$-qubit Pauli gate. Then,  whenever $\Omega$-random circuits quickly increase the weight of errors, $r$ is a good estimate of the probability that an error happens on a $\Omega$-average native gate. That is, $r  \approx \epsilon_{\Omega} \equiv \sum_{i}\Omega(\mathcal{G}_i)\epsilon_i$, where $\epsilon_i$ is the probability of an error on the $n$-qubit native gate $\mathcal{G}_i$. Later, we derive this relationship.

Because $r$ depends on the sampling distribution, they should be reported together. A similar, but hidden variability also exists in CRB -- the CRB $r$ depends on the Clifford compiler. This compiler-dependence in CRB is inconvenient, as the properties of multi-qubit Clifford compilers are difficult to control.
In contrast, because we directly choose $\Omega$, we can control how often each gate appears in the random circuits, to estimate error rates of particular interest.

\vspace{0.1cm}
\noindent
{\bf Experiments on 2 -- 5 qubits --} To demonstrate that DRB is useful and behaves correctly on current multi-qubit devices, we used it to benchmark 2 -- 5 qubit subsets of the publicly accessible IBMQX5 \cite{ibmqx,ibmqx-backend}. The IBMQX5 native gates comprise \textsc{cnot}s and arbitrary 1-qubit gates \cite{ibmqx-backend,qiskit}; we benchmarked a set of $n$-qubit gates consisting of parallel applications of all directly available \textsc{cnot}s and all 1-qubit Clifford gates.

 Fig.~\ref{Fig:IBMQX-decays} summarizes our results. Fig.~\ref{Fig:IBMQX-decays} A demonstrates that DRB was successful on 2 -- 5 qubits: an exponential decay is observed and $r$ is estimated with reasonable precision (bootstrapped $2\sigma$ uncertainties are shown). To our knowledge, this is the largest number of qubits holistically benchmarked to date, which was made possible by the streamlined nature of DRB (see Fig.~\ref{Fig:schematic}). To interpret these results it is necessary to specify the circuit sampling. Each layer was sampled as follows: with probability $p_{\textsc{cnot}}$ we uniformly choose one of the \textsc{cnot}s and add it to the sampled layer; for all $n$ or $n - 2$ remaining qubits we independently and uniformly sample a 1-qubit gate and add it to the layer. For the data in Fig.~\ref{Fig:IBMQX-decays} A, $p_{\textsc{cnot}} = 0.75$. We also implemented experiments with $p_{\textsc{cnot}} = 0.25$; see the Supplemental Material for this data and further experimental details.

Using this sampling, the average number of \textsc{cnot}s per layer is $p_{\textsc{cnot}}$, independent of $n$. Therefore $r$ would vary little with $n$ if \textsc{cnot} errors dominate, the error rates are reasonably uniform over the \textsc{cnot}s, and $n$-qubit benchmarks are predictive of benchmarks on more than $n$ qubits. Instead, the observed $r$ increases quickly with $n$. This is quantified in Fig.~\ref{Fig:IBMQX-decays} C, where we compare each observed $r$ to a prediction $r_{\text{cal}}$ obtained from the IBMQX5 CRB calibration data (1-qubit error rates from simultaneous 1-qubit CRB  \cite{ibmqx-backend,note-ibmqx,gambetta2012characterization} and \textsc{cnot} error rates from CRB on isolated pairs \cite{note-ibmqx}). These predictions are calculated both using $r \approx \epsilon_{\Omega}$ and via a DRB simulation using a crosstalk-free error model that is consistent with the calibration data. Both methods agree, confirming that the increase in $r$ with $n$ is \emph{not} due to a failure of DRB. For $n = 2$, $r_{\text{cal}}$ and $r$ are similar, demonstrating that $n$-qubit DRB and CRB are consistent. But, as $n$ increases, $r$ diverges from $r_{\text{cal}}$. This shows that the effective error rates of the 1-qubit and/or 2-qubit gates in the device change as we implement circuits over more qubits, demonstrating that $n > 2$-qubit DRB can detect errors that are not predicted by 1- and 2-qubit CRB (calibration data) or 2-qubit DRB (our data). This highlights the value of holistic benchmarking for multi-qubit devices.

Using the data from Fig.~\ref{Fig:IBMQX-decays} A ($p_{\textsc{cnot}} = 0.75$) alongside additional data with $p_{\textsc{cnot}} = 0.25$ sampling \footnote{This data was obtained after the device had been recalibrated. This is not ideal, but the error rates did not change substantially; details in the Supplemental Material.}, we can estimate the average \textsc{cnot} error rate in $n$-qubit circuits. For each $n$ and using $r \approx \sum_i\Omega(\mathcal{G}_i)\epsilon_i$, we have $\vec{r} \approx M\vec{\epsilon}$ where: $\vec{r} = (r_{0.75}, r_{0.25})$ with $r_{0.75}$ (resp., $r_{0.25}$) the $r$ obtained with $p_{\textsc{cnot}} = 0.75$ (resp., $p_{\textsc{cnot}} = 0.25$) sampling; $\vec{\epsilon} = (\epsilon_{A}, \epsilon_{B})$ with $\epsilon_A$ (resp., $\epsilon_B$) the average error rate of those $n$-qubit gates containing one \textsc{cnot} in parallel with 1-qubit gates on the other qubits (resp., $n$ parallel 1-qubit gates); $M = \frac{1}{4}(\begin{smallmatrix}3&1\\1&3\end{smallmatrix})$. Therefore, $\epsilon_A$ and $\epsilon_{B}$ can be estimated using $\vec{\epsilon} = M^{-1}\vec{r}$, and so -- by estimating the average 1-qubit gate error rate from $\epsilon_{B}$ and removing this contribution from $\epsilon_A$ -- we can estimate the mean \textsc{cnot} error rate versus $n$. Estimates are given in Fig.~\ref{Fig:IBMQX-decays} D. For two qubits, our estimate of the \textsc{cnot} error rate is similar to the prediction from the calibration data, so our methodology seems consistent with CRB techniques. In contrast, our results show that \textsc{cnot}s perform substantially worse in $n > 2$ qubit circuits than in 2-qubit circuits. This is likely due to \textsc{cnot} crosstalk, i.e., \textsc{cnot}s affect ``spectator'' qubits. 

\begin{figure}[b!]
\includegraphics[width=0.49\textwidth]{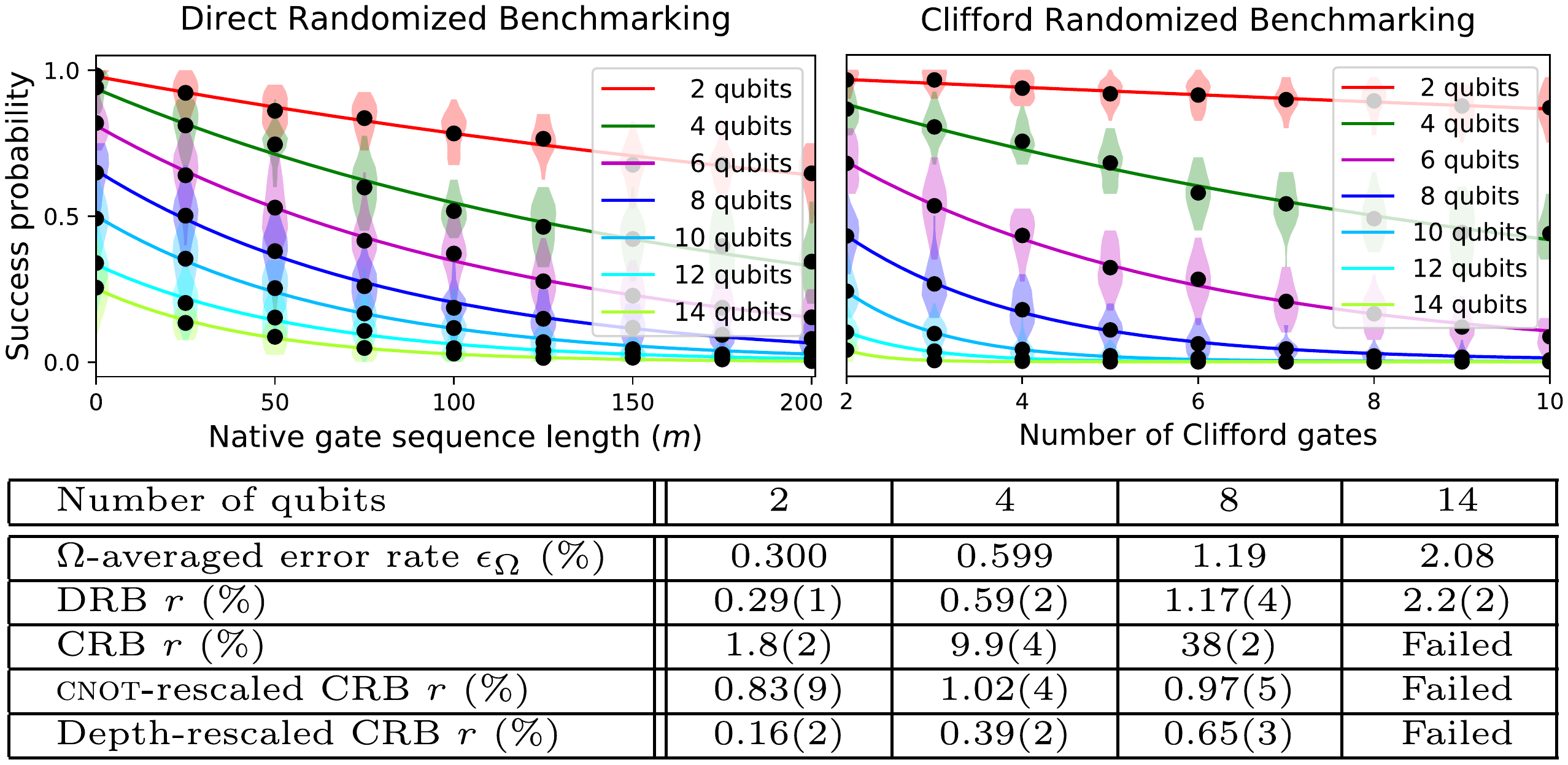}
\caption{Simulation of DRB and CRB for 2 -- 14 qubits with a simple error model. The $n$-qubit DRB error rate is $r \approx n \times 0.15  \%$, consistent with the simulated sampling-averaged native gate error rate $\epsilon_{\Omega}$.}
\label{Fig:comparing-CRB-and-DRB-varying-qubit-number-iid-dep}
\end{figure}

\vspace{0.4cm}
\noindent
{\bf DRB simulations --} We have shown that DRB works on current multi-qubit devices, and so we now demonstrate with simulations that $r \approx \epsilon_{\Omega} \equiv \sum_i\Omega(\mathcal{G}_i)\epsilon_i$. Assume $n$ qubits with native gates consisting of parallel $\textsc{cnot}$, idle $I$, Hadamard $H$ and phase $P$ gates ($P\ket{x} = i^x\ket{x}$), and all-to-all connectivity. We model gate errors by assuming that after each \textsc{cnot} (resp., 1-qubit gate) the qubits involved in the gate are independently, with probability $0.25\%$ (resp., $0.05\%$), subject to a random $\sigma_x$, $\sigma_y$ or $\sigma_z$ error. So the \textsc{cnot} error rate is $\approx 0.5\%$. We simulated DRB with a sampling distribution defined by randomly pairing up the qubits, applying a \textsc{cnot} to a pair with probability $\tilde{p}_{\textsc{cnot}} = 0.5$, and applying uniformly random 1-qubit gates ($H$, $P$ or $I$) to all qubits that do not have a \textsc{cnot} acting on them. Fig.~\ref{Fig:comparing-CRB-and-DRB-varying-qubit-number-iid-dep} shows simulated 2 -- 14 qubit DRB and CRB data. DRB has succeeded: the decay is exponential and $r \approx \epsilon_{\Omega} = 1 - (0.5\times 0.9975^2 + 0.5\times 0.9995^2)^{\frac{n}{2}} \approx n \times 0.15\%$. In contrast, the CRB $r$ grows rapidly with $n$ -- for only 4-qubit CRB $r \approx 10\%$ -- and CRB fails for $n > 12$, demonstrating that DRB can be implemented on more qubits than CRB. Moreover, the CRB error rates $r$ rescaled to $r_{\textsc{rcrb}} = 1 - (1 - r)^{1/\alpha}$ \cite{muhonen2015quantifying,barends2014superconducting,raftery2017direct,epstein2014investigating}, with $\alpha$ the average compiled Clifford circuit-depth or \textsc{cnot}-count, are not simple functions of the native gate error rates (see Fig.~\ref{Fig:comparing-CRB-and-DRB-varying-qubit-number-iid-dep}).

This example is illustrative, but simplistic. So, in the Supplemental Material we present additional simulations with large, non-uniform \textsc{cnot} error rates, and limited qubit connectivity. We also simulate the \textsc{cnot} error rate estimation method used on the IBMQX5 data, validating the technique.

\vspace{0.1cm}
\noindent
{\bf DRB theory --} We now provide a theory for DRB~of gates with Pauli-stochastic errors. DRB circuits consist of preparing a uniformly random $n$-qubit stabilizer state $\psi$, a circuit $\mathcal{U}_m = \mathcal{G}_{s_m}\cdots\mathcal{G}_{s_1}$ with $m$ layers $\mathcal{G}_{i}$ sampled according to $\Omega$, and a stabilizer measurement projecting onto $\mathcal{U}_m\ket{\psi}$. For now, assume that the stabilizer state preparation and measurement (SSPAM) are perfect. In the stochastic error model, each time $\mathcal{U}_m$ is applied there is some faulty implementation, $\tilde{\mathcal{U}}_{m} = \mathcal{P}_{s_m}\mathcal{G}_{s_m}\cdots\mathcal{P}_{s_1}\mathcal{G}_{s_1}$, with $\mathcal{P}_{s_i}$ some Pauli error or the identity. DRB aims to capture the rate that these $\mathcal{P}_{s_i}$ deviate from the identity. Because $\psi$ is a stabilizer state, the measurement will register success iff one of the following holds:
\begin{enumerate}[label=(S\arabic*)]
    \item No errors occur in $\tilde{\mathcal{U}}_{m}$, i.e., all $\mathcal{P}_{i} = \mathds{1}$.
    \item 2+ errors occur in $\tilde{\mathcal{U}}_{m}$, but when they propagate through the circuit they cancel, i.e., multiple $\mathcal{P}_i \neq \mathds{1}$ but $\tilde{\mathcal U}_m = i^k\mathcal{U}_m$ (for some $k=0,1,2,3$).
    \item 1+ errors occur in $\tilde{\mathcal{U}}_{m}$ that do not cancel, but they are nonetheless unobserved by the stabilizer measurement, i.e., $ \tilde{\mathcal{U}}_{m} \neq i^k\mathcal{U}_m$ but $\tilde{\mathcal{U}}_{m}\ket{\psi} =i^l\mathcal{U}_m\ket{\psi}$.
\end{enumerate}
The DRB average success probability $P_m$ is obtained by averaging $P(\mathcal{U}_m,\psi) = \vert\langle\psi\vert\mathcal{U}_m^{\dagger}\tilde{\mathcal{U}}_m\vert\psi\rangle\vert^2$ over the possible Pauli errors, $\psi$ and $\mathcal{U}_m$. We may then write $P_m = s_1 + (1 - s_1)(s_2 + (1 - s_2)s_3)$, where $s_{1}$ is the probability of S1, i.e., no errors, $s_{2}$ is the probability of S2 conditioned on $1+$ errors occurring, and $s_{3}$ is the probability of S3 conditioned on $1+$ errors occurring and the errors not canceling. Because $\epsilon_\Omega$ is the $\Omega$-averaged error rate per layer, $s_1 = (1 - \epsilon_{\Omega})^m$. A uniformly random stabilizer state $\psi$ is an eigenstate of \emph{any} Pauli error with probability $(2^n - 1)/(4^n - 1)$, so $s_3=(2^n - 1)/(4^n - 1)\approx2^{-n}$. This is one of the motivations for the state preparation step in DRB.

In order to understand the effect of $s_2$ on $P_m$, we consider two regimes: small $n$ ($\lesssim 3$ qubits) and not-so-small $n$ ($\gtrsim 3$). In both regimes, we expect Pauli errors to occurring at most once every several layers and to be \emph{low-weight}, with support on only a few qubits. In the not-so-small $n$ regime, errors propagating through a sequence of one- and two-qubit gates are likely to quickly increase in weight \cite{brown2015decoupling,sekino2008fast,brandao2016local} (due to the demands we made of $\Omega$ earlier). Subsequent errors are therefore very unlikely to cause error cancellation. If each layer is a uniformly random Clifford (as in uncompiled CRB), any Pauli error is randomized to one of the $4^n-1$ possible $n$-qubit Pauli errors at each step. So the probability that another error cancels with an earlier error is $\simeq 1/4^n$, implying that $s_{2} \lesssim1/4^n$. In DRB, we expect error cancellation at a rate only slightly above this. Therefore, $s_2$ contributes negligibly to $P_m$, so $P_m \approx 2^n + (1 - 2^n)(1 - \epsilon_{\Omega})^m$. This is an exponential with decay rate $\epsilon_\Omega$.  Verifying this error scrambling process for a given sampling distribution, $\Omega$, is computationally efficient in qubit number. Distributions that do $not$ scramble the errors quickly (e.g., if 2-qubit gates are rare) can yield decays that are not simple exponentials. These should be avoided.

For small $n$, the probability of cancellation ($s_2$) is \emph{not} negligible for \emph{any} distribution. But because $n$ is small, we only need a few random circuit layers of Clifford-group generators to implement approximate Clifford twirling, so $P_m$ may be computed using the resulting effective depolarizing channel. Such channels are well-known to lead to exponential decays \cite{magesan2012characterizing}. However, $s_2$ (a function of $m$) now contributes significantly to the DRB decay constant $p$, so $p \napprox 1 - \epsilon_{\Omega}$. This motivates $r =(4^n - 1) (1 - p)/4^n$, which removes the unwanted $s_2$ contribution in $1-p$. Let each layer be followed by a depolarizing map $\mathcal{D}_{\lambda}$ where $\mathcal{D}_{\lambda}[\rho] = \lambda \rho + (1 - \lambda)\mathds{1}/2^n$. Then $P_m = (1-2^{-n})\lambda^m + 2^{-n}$, but the error rate of $\mathcal{D}_{\lambda}$ is $\epsilon=(4^n - 1) (1 - \lambda)/4^n$.  Of course, in the large-$n$ limit, $\epsilon \rightarrow 1 - \lambda$.

Above, we assumed perfect SSPAM which is unrealistic. The SSPAM operations are \emph{almost} $m$-independent, and so errors in SSPAM are almost entirely absorbed into $A$ and $B$ in $P_m = A + Bp^m$, as normal in RB \cite{magesan2012characterizing,magesan2011scalable}. The only $m$-dependent impact is from correlations in the stabilizer state that is prepared and measured -- they are perfectly correlated (resp., uncorrelated) at $m = 0$ (resp., $m \to \infty$). This causes an inconsequentially small tendency to over-estimate the gate error rate -- because SSPAM contributes an error of $1-\text{avg}_i[(1-\epsilon_{i,\textsc{sspam}})^2]$ at $m=0$ but a smaller error of $1-(\text{avg}_i[1-\epsilon_{i,\textsc{sspam}}])^2$ at $m \to \infty$, where $\epsilon_{i,\textsc{sspam}}$ is the error in creating or measuring the $i^{\rm th}$ stabilizer state. 

DRB remains effective with coherent errors -- with any 1-qubit gates that generate the 1-qubit Clifford group, independently random 1-qubit gates on each qubit are  sufficient to quickly twirl coherent errors to Pauli-stochastic errors, implying that errors can only coherently combine between a few layers in a DRB circuit (in contrast to the uncontrolled coherent addition \emph{within} a compiled Clifford gate in CRB). But linking $r$ to a formal notion of gate error rate is subtler with coherent errors, in direct analogy with CRB \cite{proctor2017randomized,wallman2017randomized,carignan2018randomized}, as will be discussed in future work.

\vspace{0.1cm}
\noindent
{\bf Conclusions --} Benchmarking methods for multi-qubit systems are essential for assessing the performance of current and near-term quantum processors. But currently there are no reliable methods that can be easily and routinely applied to more than two qubits with current device performance. In this Letter we have introduced and demonstrated \emph{direct randomized benchmarking} (DRB), a method that streamlines the industry-standard Clifford randomized benchmarking (CRB) technique \cite{magesan2012characterizing,magesan2011scalable} so that it can be applied to more qubits. DRB retains the core simplicity of CRB, our protocol directly measures the quantities of most interest -- the error rates of the native gates in a device -- and it is user-configurable, allowing a variety of important error rates to be estimated. Our experimental demonstrations were on 2 -- 5 qubits, and, using a publicly accessible device \cite{ibmqx-backend,ibmqx}, set a record for the number of qubits holistically benchmarked. The tools we used are available as open-source code \cite{pygstiversion0.9.8}, and support any device connectivity. So, we anticipate that 5 -- 10+ qubits will soon be benchmarked with our protocol, providing important insights into state-of-the-art device performance. Finally, the techniques of DRB can also be applied to extend and improve the full suite of RB methods \cite{wood2017quantification,chasseur2015complete,wallman2015robust,wallman2015estimating,feng2016estimating,sheldon2016characterizing,gambetta2012characterization,harper2017estimating,chasseur2017hybrid,carignan2015characterizing,cross2016scalable,magesan2012efficient,harper2017estimating,sheldon2016characterizing}, and varied-sampling DRB provides an alternative to both interleaved CRB \cite{magesan2012efficient} and ``interleaved DRB'' for estimating individual error rates, demonstrating the broad applicability and impact of DRB.

\vspace{0.1cm}
\noindent
{\bf Acknowledgements --}
\input acknowledgement.tex
\bibliography{../../../../../library/Bibliography}
\appendix

\section*{Supplemental Material}
In this Supplemental Material, we provide futher details of the simulations, experiments, and analyses discussed in the main text to assist the reader in reproducing our results. Note that there are also ancillary files accompanying this arXiv posting, in the form of data and Python code. In the first section, we detail our implementations of the random circuit sampling used in the DRB experimental demonstration as well as in the DRB and CRB simulations. The second section provides an additional simulation with high error rates that vary from gate to gate. This example also illustrates and validates our method for estimating error rates for different sets of gates from DRB data, which we used in the main text to analyze experimental data. The final section provides additional details of our DRB experiments. This section includes the DRB decay curves obtained for the second sampling distribution used in the main text.

\subsection{Circuit sampling methods}
To implement DRB and/or CRB it is necessary to efficiently sample from the relevant class of quantum circuits. This requires efficient sampling from the set of $n$-qubit stabilizers or Cliffords, compilation of stabilizers and Cliffords into native gates, and algebraic manipulation of Clifford operations.  There are many choices to be made when implementing these steps, and we outline our methods here. Note that our circuit sampling code has been integrated into the open-source \texttt{pyGSTi} Python package \cite{pygstiversion0.9.8}, and so it is publicly available.  

Given some sampling distribution $\Omega$, the tasks required to sample an $n$-qubit DRB circuit at sequence length $m$ are: 
\begin{enumerate}
\item [D1.] Uniformly sample a stabilizer state $\ket\psi$.
\item [D2.] Generate a circuit of native gates that creates the stabilizer state $\ket\psi$ from $\ket 0^n$.
\item [D3.] Sample a depth-$m$ circuit, $\mathcal{U}_m$, with each layer sampled according to the distribution $\Omega$ over the native gates.
\item [D4.] Calculate the stabilizer state $\ket\varphi = \mathcal{U}_m\ket\psi$.
\item [D5.]  Generate a circuit that maps the stabilizer state $\ket\varphi $ to some computational basis state $\ket s$.
\end{enumerate}

We implement task D3 in the obvious way -- sampling circuit layers according to a user-specified distribution is computationally simple with all the distributions that we used for our simulations and experiments. Before discussing how we achieved the remaining tasks in DRB circuit sampling, we overview the tasks required to sample CRB circuits.

The tasks required to sample an $n$-qubit CRB circuit, consisting of $m$ random Clifford gates followed by an inversion Clifford, are:
\begin{enumerate}
\item [C1.] Uniformly sample $m$ Clifford gates.
\item [C2.] Calculate the unique Clifford gate that inverts this sequence of $m$ Clifford gates.
\item [C3.] Compile each of these $m+1$ Clifford gates into a circuit over the native gates.
\end{enumerate}
 
To implement all these steps in generating DRB and CRB circuits (except step D3) we used the symplectic matrix representation of Clifford gates \cite{dehaene2003clifford,hostens2005stabilizer}. Specifically, a Clifford gate on $n$-qubits is represented by a unique $2n \times 2n$ symplectic matrix over $[0,1]$ alongside a unique length $2n$ vector over $[0,1,2,3]$ \cite{dehaene2003clifford,hostens2005stabilizer}. 

The first task for generating a CRB circuit (C1) is uniform sampling over the $n$-qubit Clifford group. To do this we used the method and Python code of Koenig and Smolin \cite{koenig2014efficiently}, which uniformly samples a $2n \times 2n$ symplectic matrix over [0,1] (we integrated this code into the \texttt{pyGSTi} Python package \cite{pygstiversion0.9.8}). A symplectic matrix only partially defines a Clifford gate (see above), so we pair the sampled matrix $S$ with a random $2n$ vector over $[0,1,2,3]$, chosen uniformly from all those vectors $v \in [0,1,2,3]^{2n}$ that together with $S$ define a valid Clifford gate (not all $S$ and $v$ pairs define a valid Clifford -- given any $S$, half of the possible vectors define a valid Clifford \cite{dehaene2003clifford,hostens2005stabilizer}).

The first task for generating a DRB circuit (D1) is uniform sampling over the set of $n$-qubit stabilizer states. The action of an $n$-qubit Clifford gate $\mathcal{C}$ on $\ket 0^n$ defines a stabilizer state $\mathcal{C}\ket{0}^n$. So to uniformly sample a stabilizer state we uniformly sample an $n$-qubit Clifford gate $\mathcal{C}$, using the method above, and take the sampled stabilizer state to be $\mathcal{C}\ket{0}^n$. There are more computationally efficient ways to sample random stabilizer states, but this method is sufficient for our purposes.

Tasks C2 and D4 require computing the Clifford operation implemented by a sequence of Clifford gates -- this is simple in the symplectic representation, using matrix multiplication of $2n \times 2n$ matrices. Explicit formulas are given in Ref.~\cite{hostens2005stabilizer}. Task C2 also requires finding the inverse of a Clifford gate, and this is also implemented using the formulas in Ref.~\cite{hostens2005stabilizer}.

The remaining tasks for generating DRB and CRB circuits consist of circuit compilation. To implement C3 we need a method to compile any Clifford gate into a circuit over the native gates. To do this we use a Gaussian-elimination-based algorithm that is similar to that presented in Ref.~\cite{hostens2005stabilizer}. This uses $O(n^2)$ 1- and 2-qubit gates, and so it does not have the asymptotical optimal scaling of $O(n^2/\log n)$ 1- and 2-qubit gates \cite{aaronson2004improved,patel2003efficient}. Our algorithm incorporates an element of randomization in order to find lower \textsc{cnot}-count compilations, that is equivalent to repeating a similar algorithm to that in Ref.~\cite{hostens2005stabilizer} with the (arbitrary) qubit labelling, from 1 to $n$, randomized. As our code is open-source, the exact algorithm we use can be found at Ref.~\cite{pygstiversion0.9.8}.

To implement the stabilizer-state-generation compilation in D3 and D5 we use a method based on Ref.~\cite{aaronson2004improved}. In Ref.~\cite{aaronson2004improved} it is shown how to generate any stabilizer state using a circuit consisting of a layer of 1-qubit gates, a \textsc{cnot} circuit, and then another layer of 1-qubit gates. Using an asymptotically optimal \textsc{cnot} circuit compiler this would use $O(n^2/\log n)$ 1- and 2-qubit gates \cite{patel2003efficient}, but our \textsc{cnot} circuit compiler -- like our general Clifford gate compiler -- is based on Gaussian elimination, and so it uses $O(n^2)$ \textsc{cnot} gates. As such, it is directly comparable to our Clifford gate compiler, meaning that our comparisons between DRB and CRB are reasonably fair (it is not clear how to make sure that the comparison is completely fair without access to optimal stabilizer state and Clifford gate compilers). Finally, note that our stabilizer state compiler also uses an equivalent form of randomization to our Clifford gate compiler, as well as a heuristic method for reducing the \textsc{swap}-gate-like over-head associated with limited qubit connectivity, that is based on optimizing the order in which qubits are ``eliminated'' in the Gaussian elimination. The exact compilation methods can be found in the open-source code at Ref.~\cite{pygstiversion0.9.8}.

\subsection{Simulations}
\label{app:sims}
In the first part of this section we overview our simulation methods. In the second part of this section we present an additional simulated example of DRB and CRB. This simulation is on five qubits with realistic connectivity, high error rates that vary from gate to gate, and realistic state preparation and measurement error rates.

\subsubsection{Methods}
To simulate DRB and CRB we require a circuit simulator that can handle the error models we consider. All the errors models herein consist of Pauli-stochastic errors on the gates (and the measurements, in the case of the simulation later in this section). For this reason, all the simulations presented herein use an efficient-in-$n$ simulator that samples Pauli errors according to the model-specified error probability statistics (where the probability of an error is generally dependent on the gate that is applied), and propagates them through the Clifford circuit using the symplectic representation of Clifford gates (see the previous section). I.e., this is a basic Pauli errors in Clifford circuits simulator that uses an unravelling of the stochastic error maps. Python Jupyter notebooks, that we used to run our simulations, the raw data we obtained, and the data analysis, can be found in auxiliary files accompanying this arXiv posting

Although all the DRB and CRB simulations presented herein used Pauli-stochastic error models, pure state and density matrix circuit simulators are also available within \texttt{pyGSTi}, and so multi-qubit DRB and CRB simulations with more general error models can be easily implemented for moderate numbers of qubits. Indeed, we have simulated DRB with a wide range of error models -- including purely coherent errors -- and have found DRB to be broadly reliable. Additional DRB simulations will be presented in future papers.

\subsubsection*{An additional simulated example of DRB and CRB}
\begin{figure}
\includegraphics[width=0.49\textwidth]{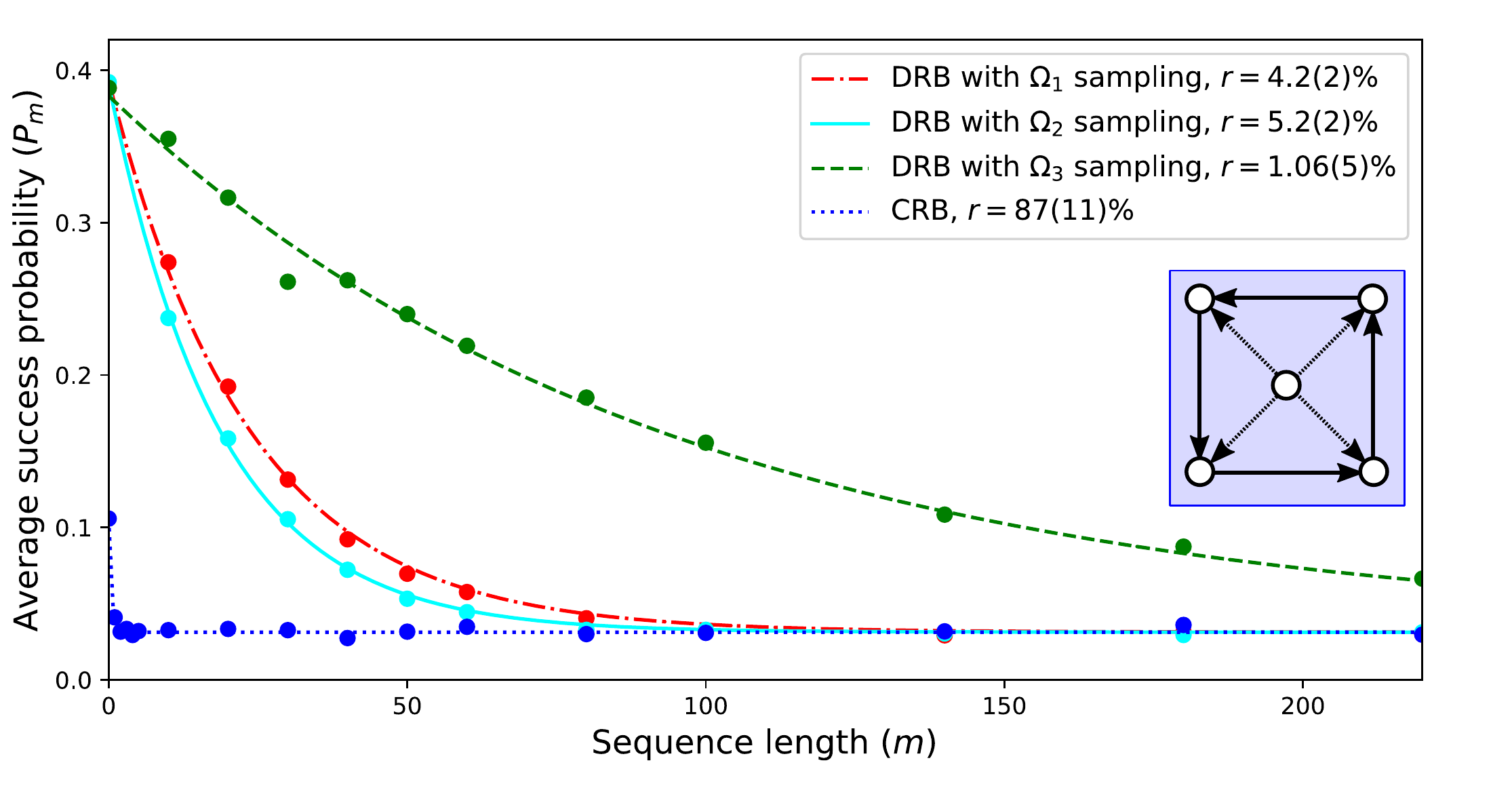}
\caption{Simulated CRB and DRB on 5 qubits with large and strongly gate-dependent stochastic errors. Inset: The connectivity of the five qubits. The arrows denote \textsc{cnot} gates and point from the control to the target qubit. Main figure: Average success probabilities $P_m$ (points) and fits to $P_m = A + Bp^m$ (curves). DRB has been simulated with three different sampling distributions, $\Omega_{1}$, $\Omega_2$ and $\Omega_3$. The $\Omega_1$ and $\Omega_2$ distributions have the same average \textsc{cnot} count per random circuit layer, which is 0.75. The difference is that $\Omega_1$ (resp., $\Omega_2$) is weighted towards a subset of the \textsc{cnot}s which happen to have lower (resp., higher) error rates. The $\Omega_3$ distribution is equally likely to select any of the \textsc{cnot}s in the device, and the probability that any given circuit layer contains a \textsc{cnot} is $0.10$.}
\label{fig:DRB-decays-high-error-rate}
\end{figure}

The simulated example of DRB and CRB presented in the main text was illustrative, but the error model was not particularly representative of current multi-qubit devices. Here we present additional simulations with large, non-uniform \textsc{cnot} error rates, and limited qubit connectivity. Consider five qubits with the connectivity shown in the inset of Fig.~\ref{fig:DRB-decays-high-error-rate}, which has a connectivity graph that is typical of small superconducting devices \cite{ibmqx,ibmqx-backend}. As in the main text, we consider a native gate set that permits simultaneous application of a \textsc{cnot} gate on any two connected qubits and single-qubit gates (idle, Hadamard and phase gates) on any qubits. The \textsc{cnot} gates are taken to have fixed directionality, with the directions shown in the inset of Fig.~\ref{fig:DRB-decays-high-error-rate}. There are 8 \textsc{cnot} gates in this device: 4 on the outer ring of 4 qubits, and 4 from the central qubit to each of the outer qubits. 

Consider an as-implemented device whereby:
\begin{itemize}
\item Each 1-qubit gate is subject to a stochastic Pauli error with probability $0.1\%$. If an error does occur, it is uniformly likely to be a $\sigma_x$, $\sigma_y$ or $\sigma_z$ error.
 
\item \textsc{cnot} gates for which both qubits are contained in the outer ring have a $4\%$ total error rate. Specifically, both the control and target qubits are independently subject to a uniformly random Pauli stochastic error with a probability of error on each qubit of $1-(1-0.04)^{1/2} \approx 0.02$.

\item \textsc{cnot} gates connecting the central qubit and the outer ring qubits all have an $8\%$ total error rate. Specifically, the central qubit is, with a probability of $4\%$, subject to a uniformly random Pauli stochastic error. Additionally,  \emph{all} of the qubits in the outer ring are independently subject to a uniformly random Pauli stochastic error with a probability of error on each qubit of $1-[(1-0.08)/(1-0.04)]^{1/4} \approx 0.01$. Thus, these particular \textsc{cnot} gates have crosstalk errors.

\item There are 2\% measurement errors on every qubit. Specifically, each qubit is independently subject to a $\sigma_x$ bit-flip error with probability $2\%$ immediately before it is perfectly measured. Therefore, the total measurement error is $\approx 10\%$.
\end{itemize}

We simulated DRB and CRB with this error model -- the resultant data is shown in Fig.~\ref{fig:DRB-decays-high-error-rate}. There are three sets of data for DRB, as we simulated DRB with three different sampling distributions, which we define below. For all three distributions, the DRB decay is an exponential and reasonably high precision error rates are extracted. In contrast, the CRB data is essentially unusable, because the error rate is nearly as high as it could possibly be, and the estimate of $r$ is very imprecise. Note that we have used physically practical sampling parameter: 50 circuits have been sampled at each length, there are 100 counts for each circuit, and for DRB (resp., CRB) data has been obtained at 12 (resp., 17) distinct lengths.

The DRB error rates are perhaps not immediately intuitive, because with an error model with strongly gate-dependent error rates the DRB $r$ will strongly depend on the sampling distribution, and the sampling has not yet been specified. The three different sampling distributions we used are all particular cases of the following parameterized distribution $\Omega[\vec{v} = (v_1,v_2,v_3)]$, defined by the following algorithm to sample a circuit layer: 
\begin{enumerate}
\item Draw $S$ from $[1,2,3]$ according to the distribution $\vec{v} = (v_1,v_2,v_3)$ with $v_1+v_2+v_3 = 1$.
\item If $S=1$ then independently and uniformly at random sample a 1-qubit gate to apply to each qubit.
\item If $S=2$ then uniformly sample a \textsc{cnot} gate on the outer ring, and then independently and uniformly at random sample a 1-qubit gate to apply to each remaining qubit.
\item If $S=3$ then uniformly sample a \textsc{cnot} gate from the central qubit to a qubit on the ring, and then independently and uniformly at random sample a 1-qubit gate to apply to each remaining qubit.
\end{enumerate}

The three particular sampling distribution we simulated are given by $\Omega_{k} = \Omega[\vec{v}_k]$ for $k=1,\,2,\,3$ where
\begin{align}
\vec{v}_1= \frac{1}{4} \begin{pmatrix} 1 \\ 2\\ 1 \end{pmatrix}, \hspace{0.5cm}
\vec{v}_2= \frac{1}{4} \begin{pmatrix} 1 \\ 1\\ 2 \end{pmatrix}, \hspace{0.5cm}
\vec{v}_3= \frac{1}{20} \begin{pmatrix} 18 \\ 1\\ 1 \end{pmatrix}.
\end{align}
The exact values chosen have no special significance, but the general form has been chosen so that we can extract precise estimates of the error rates of different gates, as we show below, and so that we would expect DRB to be reliable -- error are ``scrambled'' reasonably quickly with all three distributions.

In the main text we claim that, for stochastic errors and sampling distributions that quickly scramble errors, $r \approx \sum_i\Omega(\mathcal{G}_i) \epsilon_i$ where $\epsilon_i$ is the error rate for gate $\mathcal{G}_i$ and $\Omega$ is the sampling distribution. Therefore, because of the sampling we have chosen, the three different DRB error rates we have estimated are different weighted sums of three terms, $\epsilon_1$, $\epsilon_2$ and $\epsilon_3$, defined below.
\begin{itemize}
\item
Let $\epsilon_1$ denote the average error rate over all $3^5$ of the 5-qubit native gate obtained from the tensor product of 1-qubit gates only. This is the average error rate of the gates that are uniformly sampled from with probability $v_1$. With this error model, $\epsilon_1 = 1-(1-0.001)^5 \approx 0.005$.

\item Let $\epsilon_2$ denote the average error rate over all $3^3 \times 4$ of the 5-qubit native gates obtained from any \textsc{cnot} gate between qubits on the outer ring in parallel with 1-qubit gates on the other qubits. This is the average error rate of the gates that are uniformly sampled from with probability $v_2$. With this error model, $\epsilon_2 = 1-(1-0.04)(1-0.001)^3 \approx 0.043$.

\item Let $\epsilon_3$ denote the average error rate over all $3^3 \times 4$ of the 5-qubit native gates obtained from any \textsc{cnot} gate between the central qubit and a qubit on the outer ring in parallel with 1-qubit gates on the other qubits. This is the average error rate of the gates that are uniformly sampled from with probability $v_3$. With this error model
\begin{align*}
\,\,\,\,\,\,\epsilon_3 &= (1-0.04)(1-\eta)^4(1-0.001)^3 - O(0.001 \eta ) ,\\&\approx 0.083,
\end{align*}
where $\eta =1-(1-0.04)^{1/4} \approx 0.0125$. The source of the $O(0.001 \eta )$ term is that, for these 5-qubit gates, there are two independent sources of error on three of the outer qubits and, if these errors both happen and cancel, then the overall effect is no error. This happens with a small enough probability to be irrelevant.
\end{itemize}

Let $r(\Omega_k)$ denote the DRB error rate under the sampling distribution $\Omega_k$ and let
\begin{equation}
\vec{\epsilon} = \begin{pmatrix} \epsilon_1 \\ \epsilon_2 \\ \epsilon_3 \end{pmatrix}.
\end{equation}
Under the sampler $\Omega_k$ we uniformly sample a gate from a class of gates with an average error rate of $\epsilon_j$ with probability $v_{k,j}$. Therefore, according to our theory of the DRB error rate $r$, we have that 
\begin{equation}
r(\Omega_k) \approx \vec{v}_k \cdot \vec{\epsilon}.
\label{eq:r-as-specific-sum-of-espilon}
\end{equation}
By evaluating the RHS of this equation we obtain $\vec{v}_1 \cdot \vec{\epsilon} = 0.0434$, $ \vec{v}_2 \cdot \vec{\epsilon} = 0.0533$ and $ \vec{v}_3 \cdot \vec{\epsilon} = 0.0108$, to three significant figures. These predicted values for $r$ are within a standard deviation of the $r$ values that are observed in the simulations, which are shown in the legend of Fig.~\ref{fig:DRB-decays-high-error-rate}.

 Eq.~\eqref{eq:r-as-specific-sum-of-espilon} can be re-expressed as
\begin{equation}
\begin{pmatrix} 
\frac{18}{20} & \frac{1}{20} & \frac{1}{20} \\ 
 \frac{1}{4} & \frac{1}{4} & \frac{1}{2} \\
 \frac{1}{4} & \frac{1}{2} & \frac{1}{4} \\
 \end{pmatrix} \begin{pmatrix}\epsilon_1 \\ \epsilon_2 \\ \epsilon_3 \end{pmatrix}=   \begin{pmatrix}r(\Omega_1) \\ r(\Omega_2) \\ r(\Omega_3) \end{pmatrix}.
\end{equation}
So, to estimate $\epsilon_1$, $\epsilon_2$ and $\epsilon_3$, we invert the matrix on the left-hand-side of this equation, and apply this to our estimated vector of DRB error rates. The obtained estimates are shown in the top half of Table~\ref{Table:individual-error-rates-from-DRB}, along with an estimate of $(\epsilon_2 + \epsilon_3)/2$, which is of interest because it quantifies average \textsc{cnot} performance. The uncertainties have been calculated from the uncertainties in the DRB error rates and standard linear propagation of uncertainty. All the estimates are sufficiently precise for most practical purposes, and all the estimates are within 1 standard deviation of the true values of these error rates.

\begin{table}
\begin{tabular}{ | c |c | c | c |   }
\hline
Error rate & Estimate  & True value  & Relative Error  \\  
 \hline\hline		
$\epsilon_1$ & $0.0050(6)$  & 0.0050  & $-0.00003$ \\
\hline
$\frac{1}{2}(\epsilon_2 + \epsilon_3)$ & $ 0.061(2)$  & 0.063 & -0.03  \\
\hline
$\epsilon_2$  & $ 0.040(5) $ & 0.043 & -0.08 \\
  \hline  
$\epsilon_3$  & $ 0.082(7) $  & 0.083 & -0.005 \\
  \hline  
  \hline  
$\epsilon_{\text{local}}$  & $0.0010(1)$ & 0.001 & $-0.00003$ \\
  \hline  
$\epsilon_{\textsc{cnot}}$  & $0.058(2)$  & 0.06 & $-0.03$ \\
  \hline  
$\epsilon_{\textsc{r-cnot}}$  & $ 0.037(5)$  & 0.04 & $-0.08$  \\
  \hline  
$\epsilon_{\textsc{c-cnot}}$  &$0.080(7) $ & 0.08 & $-0.005$  \\
  \hline  
\end{tabular}
\caption{Estimates of a variety of error rates, for the 5-qubit device of Fig.~\ref{fig:DRB-decays-high-error-rate}, from simulated DRB. These estimates have been extracted from three simulated DRB experiments that use three different sampling distributions. The $\epsilon_1$, $\epsilon_2$ and $\epsilon_3$ are gate-averaged error rates for distinct classes of 5-qubit gates: $\epsilon_1$ is the gate-averaged error rate over all 5-qubit gates composed from tensor products of 1-qubit gates; $\epsilon_2$ is the gate-averaged error rate over all 5-qubit gates composed from a \textsc{cnot} on the outer ring in the device (see Fig.~\ref{fig:DRB-decays-high-error-rate}) in parallel with 1-qubit gates on the other qubits;  $\epsilon_3$  is the gate-averaged error rate over all 5-qubit gates composed from a \textsc{cnot} from the central qubit to a qubit on the outer ring (see Fig.~\ref{fig:DRB-decays-high-error-rate}), in parallel with 1-qubit gates on the other qubits. The $\epsilon_{\text{local}}$, $\epsilon_{\textsc{r}-\textsc{cnot}}$, $\epsilon_{\textsc{c}-\textsc{cnot}}$ and $\epsilon_{\textsc{cnot}}$ error rates are gate-averaged error rates of 1- and 2-qubit building-block gates: $\epsilon_{\text{local}}$ is the average over qubits and gates of the 1-qubit gate error rate;  $\epsilon_{\textsc{r}-\textsc{cnot}}$ is the average error rate of the 4 \textsc{cnot} gates on the outer ring; $\epsilon_{\textsc{c}-\textsc{cnot}}$ is the average error rate of the 4 \textsc{cnot} gates from the central qubit to the outer ring; $\epsilon_{\textsc{cnot}}$ is the average error rate of all  8 \textsc{cnot} gates. The relative error is defined as $(\epsilon_{\text{estimate}} -\epsilon_{\text{true}})/\epsilon_{\text{true}}$ where $\epsilon_{\text{estimate}}$ and $\epsilon_{\text{true}}$ are the estimated and true values of an error rate, respectively. Note that the true error rates have been rounded to 2 significant figures in the upper half of the table.}
\label{Table:individual-error-rates-from-DRB}
\end{table}

To estimate average \textsc{cnot} error rates,  we assume that the (uniform) average error rate of the 1-qubit gates is the same on all qubits, with this error rate denoted $\epsilon_{\text{local}}$. In this error model this approximation holds exactly, but in practice it normally would not. Under this approximation, we have that
\begin{align}
\epsilon_1  &=1 - (1-\epsilon_{\text{local}})^5, \\
\epsilon_2  &=1 - (1-\epsilon_{\text{local}})^3\left(1-\epsilon_{\textsc{r}-\textsc{cnot}}\right), \\
\epsilon_3  &=1 - (1-\epsilon_{\text{local}})^3\left(1-\epsilon_{\textsc{c}-\textsc{cnot}}\right),
\end{align}
where $\epsilon_{\textsc{r}-\textsc{cnot}}$ is the average error rate of the 4 \textsc{cnot} gates on the outer ring, and $\epsilon_{\textsc{c}-\textsc{cnot}}$ is the average error rate of the 4 \textsc{cnot} gates from the central qubit to the outer ring. 

From our estimates of $\epsilon_1$, $\epsilon_2$ and $\epsilon_3$ it is then simple to estimate $\epsilon_{\text{local}}$, $\epsilon_{\textsc{r}-\textsc{cnot}}$ and  $\epsilon_{\textsc{c}-\textsc{cnot}}$ and $\epsilon_{\textsc{cnot}}=  (\epsilon_{\textsc{r}-\textsc{cnot}}+\epsilon_{\textsc{c}-\textsc{cnot}})/2$, with this latter quantity denoting the average error rate of all 8 \textsc{cnot}s. Confidence intervals on these estimates can be obtained by a standard bootstrap. The results are shown in the lower half of Table~\ref{Table:individual-error-rates-from-DRB}.

\subsection{Experiments}
\label{app:exp}
This section contains some additional details on, and data from, our 2 -- 5 qubit DRB experiments on IBMQX5. Each individual DRB experiment used 28 circuits at each DRB length, lengths ranging from $m=0$ to $m=30$ in increments of 5, and 1024 counts per circuit. The circuits were sampled using code that has now been integrated into the open-source \texttt{pyGSTi} Python package \cite{pygstiversion0.9.8}, and using the methods described earlier in this Supplemental Material. All of the data -- including the DRB circuits used, the DRB data, and the IBMQX5 calibration data -- and Python Jupyter notebooks containing our analysis, can be found in ancillary files accompanying this arXiv posting

\begin{figure}[h!]
\includegraphics[width=0.49\textwidth]{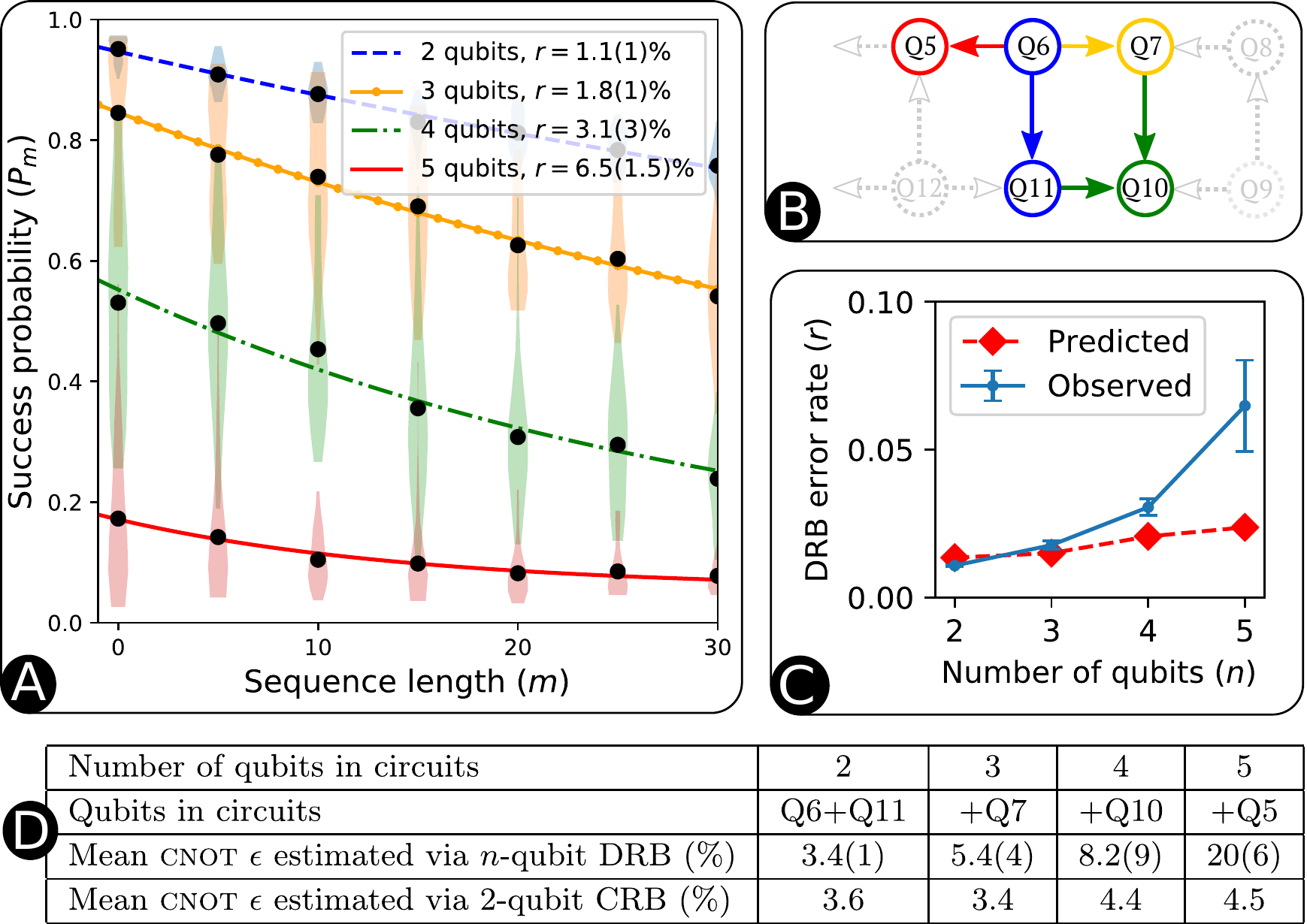}
\caption{Further 2 -- 5 qubit DRB experiments on IBMQX5. The difference from the data presented in the main text is that for that data the circuit sampling resulted in $p_{\textsc{cnot}}=0.75$ \textsc{cnot}s per layer on average, and here the circuit sampling results in $p_{\textsc{cnot}}=0.25$ \textsc{cnot}s per layer on average. As we would expect, the DRB error rate is significantly lower in this case than with the $p=0.75$ \textsc{cnot} sampling discussed in the main text. {\bf A.} Success probability decays. The points are average success probabilities $P_m$, and the violin plots show the distributions of the success probabilities at each length over circuits (there are $k_m=28$ circuit per length). The curves are obtained from fitting to $P_m = A + Bp^m$ and $r = (4^n-1)(1-p)/4^n$. {\bf B.} A schematic of IBMQX5. The colors match those in A and correspond to the additional qubits/\textsc{cnot}s added from $n \to n+1$ qubit DRB. {\bf C.} Observed $r$ versus $n$, and predictions from 1- and 2-qubit CRB calibration data. {\bf D.} Estimates of the average \textsc{cnot} error rate in $n$-qubit circuits, obtained by comparing the data in A with additional DRB data that used circuits with fewer \textsc{cnot}s per layer.}
\label{Fig:IBMQX-decays-2}
\end{figure}

As discussed in the main text, we implemented two sets of DRB experiments on IBMQX5, with different random circuit sampling. For both sets of data, an $n$-qubit circuit layer was sampled using the following algorithm:
\begin{enumerate}
\item Sample a random variable $X \in [0,1]$ with the probability that $X = 1 $ given by $p_{\textsc{cnot}}$. That is, flip a coin with bias $p_{\textsc{cnot}}$.
\item If $X = 1$:
\begin{enumerate}
\item Uniformly at random choose one of the \textsc{cnot} gates from the benchmarked $n$-qubit subset of IBMQX5 and add it to the sampled layer. There are between 1 and 5 possible \textsc{cnot} gates, depending on the number of qubits benchmarked, as shown in Fig.~\ref{Fig:IBMQX-decays-2} B. 
\item For each of the $n-2$ qubits that the sampled \textsc{cnot} gate does not act on, independently and uniformly sample a 1-qubit Clifford gate to apply to that qubit in this layer.
\end{enumerate}
\item If $X =  0$, for each of the $n$ qubits, independently and uniformly sample a 1-qubit Clifford gate to apply to that qubit in this layer.
\end{enumerate}

We implemented 2 -- 5 qubit DRB with $p_{\textsc{cnot}}=0.75$ and $p_{\textsc{cnot}}=0.25$. In the main text we presented the data for $p_{\textsc{cnot}}=0.75$. In Fig.~\ref{Fig:IBMQX-decays-2} the data for $p_{\textsc{cnot}}=0.25$ is presented. The difference between these results and those in the main text is that the DRB error rates are significantly lower in this case. Specifically, they are smaller by a multiplicative factor of between 1/2 and 1/3. With $p_{\textsc{cnot}}=0.25$ sampling there is, in expectation, $1/3$ as many \textsc{cnot} gates in the central random DRB circuit than with $p_{\textsc{cnot}}=0.75$ sampling. So, this reduction in the value of $r$, between $p_{\textsc{cnot}}=0.75$ and $p_{\textsc{cnot}}=0.25$ sampling, is inline with what we would expect if \textsc{cnot} error rates dominate but the 1-qubit error rate is non-zero. This is quantified in the main text, where we used both sets of DRB data together to estimate the average \textsc{cnot} error rates, and these results are also summarized in Fig.~\ref{Fig:IBMQX-decays-2} D.

IBMQX5 was recalibrated between when we obtained the data with $p_{\textsc{cnot}}=0.25$ and $p_{\textsc{cnot}}=0.75$. However, the gate error rates in the relevant subset of IBMQX5 -- as reported by the IBMQX5 calibration data -- only changed slightly between the times when the two sets of data were obtained. Specifically, IBMQX5 calibration data average \textsc{cnot} error rates over the 2 -- 5 qubits benchmarked varied by no more than $\approx 0.2\%$ between the two datasets, with the value of this average \textsc{cnot} error rate ranging from $\approx 3.6\%$ to $\approx 4.5\%$ from $n=2$ up to $n=5$. E.g., for $n=2$ the error rate of the one relevant \textsc{cnot} gate changed from $\approx 3.7\%$ to $\approx 3.5\%$ from the data with $p_{\textsc{cnot}}=0.25$ sampling to the data with $p_{\textsc{cnot}}=0.75$ sampling, and the change in average \textsc{cnot} error rate was smaller for all other $n$. The recalibration between data sets is not ideal given that we compare the results from the two datasets to estimate average \textsc{cnot} error rates (see the main text), but we do not think this is particularly problematic given that our experimental results are intended as a demonstration of the utility of DRB, not as a comprehensive benchmarking of IBMQX5. The raw calibration data can be found in ancillary files accompanying this arXiv posting

In the main text we compared our $n$-qubit DRB results, obtained with $p_{\textsc{cnot}}=0.75$ sampling, with predictions of the $n$-qubit DRB error rate based on the IBMQX5 calibration data (and in Fig.~\ref{Fig:IBMQX-decays-2} C the same comparison is made for the $p_{\textsc{cnot}}=0.25$ data). Here, we explain how these predictions were obtained. The IBMQX5 calibration data consists of a readout and a 1-qubit gate error rate for each qubit, and an error rate for each \textsc{cnot} gate. To our knowledge, the 1-qubit gate and \textsc{cnot} error rates are obtained by simulatenous 1-qubit CRB \cite{gambetta2012characterization,ibmqx-backend} and 2-qubit CRB on isolated pairs, respectively -- where in the latter case the 2-qubit CRB $r$ is rescaled by the average number of \textsc{cnot} gates in a compiled 2-qubit Clifford gate, after the estimated contribution of the 1-qubit gates has been removed \cite{note-ibmqx}. By assuming that these error rates are predictive of $n$-qubit gates obtained by parallel applications of these 1- and 2-qubit gates, these error rates allow us to predict the DRB error rate via $r \approx \epsilon_{\Omega}$, where $\epsilon_{\Omega}$ is the $\Omega$-averaged error rate of the $n$-qubit native gates. These are the $r_{\text{cal}}$ predictions presented in the main text.

This method for predicting the DRB error rate from CRB calibration data relies on our theory for DRB that $r \approx \epsilon_{\Omega}$ (see the main text). So there are two reasons that the observed DRB $r$ could deviate from the prediction that we derived from the 1- and 2-qubit CRB data:
\begin{enumerate}
\item The gates in IBMQX5 perform worse in circuits over more qubits. That is, even ideal characterization or benchmarking of isolated pairs of qubits would not predict the behaviour of $n > 2$ qubit circuits.
\item Our theory for DRB is wrong, and the deviation of the observed $r$ and predicted $r$ is due to a failing in DRB. E.g., the results of $n$-qubit CRB or full $n$-qubit tomography, if it was feasable, would be consist with predictions from the 1- and 2-qubit CRB data.
\end{enumerate}

In the main text we claim that (1) is the reason for the observed discrepancy between predicted and observed $r$. But, as DRB is a new methodology, we need evidence to rule out (2). To obtain this evidence, we simulated DRB using a crosstalk-free error model of IBMQX5 that is consistent with the IBMQX5 calibration data. Specifically, we modelled each gate as a perfect gate followed by independent, uniform depolarization on the qubits that the gate should act on, with the depolarization rate fixed to be consistent with the calibration data. We modelled the measurements as perfect except that they are preceeded by a bit-flip error with a probability that is consistent with the calibration data. We simulated DRB with this model using (a) exactly the same DRB circuits as used in the experiments, (b) new, independently sampled DRB circuits. Both sets of simulations result in predictions for the DRB $r$ that are consistent with the calculation based on $r \approx \epsilon_{\Omega}$. The results of these simulations, and Python Jupyter notebooks that can be used to run new simulations based on the IBMQX5 calibration data, can be found in ancillary files accompanying this arXiv posting. Note that, in addition to being good evidence for discounting possibility (2) above, these simulation results are also further general evidence in support of our theory of DRB.

\end{document}

%% file: acknowledgement.tex
TJP thanks Scott Aaronson for helpful correspondence on how to efficiently generate random stabilizer states. KR thanks Jay Gambetta, Diego Moreda and Ali Javadi for QISKit support. We acknowledge use of the IBM Q Experience for this work. This paper describes objective technical results and analysis. Any subjective views or opinions that might be expressed in the paper do not necessarily represent the views of the U.S. Department of Energy or the United States Government. Sandia National Laboratories is a multimission laboratory managed and operated by National Technology \& Engineering Solutions of Sandia, LLC, a wholly owned subsidiary of Honeywell International Inc., for the U.S. Department of Energy's National Nuclear Security Administration under contract DE-NA0003525. This research was funded, in part, by the Office of the Director of National Intelligence (ODNI), Intelligence Advanced Research Projects Activity (IARPA). All statements of fact, opinion or conclusions contained herein are those of the authors and should not be construed as representing the official views or policies of IARPA, the ODNI, or the U.S. Government.